\pdfoutput=1
\documentclass[nofootinbib,twocolumn,showpacs,amssymb,floatfix,deluxe,aps]{revtex4}
\usepackage [latin1]{inputenc}
\usepackage{graphicx}
\usepackage{dcolumn}
\usepackage{bm}
\usepackage{epsfig}
\usepackage[english]{babel}
\usepackage{amsmath}
\usepackage{amssymb}

\usepackage{verbatim}
\usepackage{alltt}

\def\({\left(}
\def\){\right)}

\begin{document}

\preprint{}
\title{
A revamped understanding of Cosmic Rays and Gamma-Ray Bursts} 

\author{A. De R\'ujula${}^{a,b}$}
\affiliation{  \vspace{3mm}
${}^a$Instituto de F\'isica Te\'orica (UAM/CSIC), Univ. Aut\'onoma de Madrid, Spain;\\
${}^c$Theory Division, CERN, CH 1211 Geneva 23, Switzerland
}

\date{\today}

\begin{abstract}
Interesting data on Gamma Ray Burts (GRBs) and Cosmic Rays (CRs) have recently been made public. 
GRB221009A has a record ``peak energy". The CR electron spectrum has been measured
to unprecedented high energies and exhibits a ``knee" akin to the ones in all-particle or
individual-element CR nuclei. IceCube has not seen high-energy neutrinos associated
with GRBs. AMS has published a CR positron spectrum conducive to much speculation.
We examine these data in the light of the ``CannonBall Model" of GRBs and CRs, 
in which they are intimately related and which they do strongly validate. 
\end{abstract}


\maketitle

{\bf{NOTE}}
A fraction of this paper elaborates results by S. Dado and A. Dar
in \cite{LastDD}.

\vspace{-.3cm}
\section{Introduction}

Amongst the many critical remarks concerning the Cannon Ball (CB) model
the most scholarly one may have been ``This is almost
Baron Munchausen" \cite{Referee}, presumably
meaning a lengthy and surprising list of fabrications. But, as I shall show,
what is really surprisingly lengthy is the list of correct predictions of the model.

The CB model \cite{CBoldies} describes long-duration gamma-ray bursts 
(referred in this paper simply as
GRBs), X-rays flashes (XRFs), short-duration GRBs originating in neutron-star mergers
\cite{Goodman2,Meszaros},
 and non-solar cosmic rays
(called here simply CRs), as well as various other astrophysical phenomena.
Early summaries can be found in \cite{DD2004}, \cite{DD2008},
the second of which will be referred to as DD2008.

Details of the model and some of its predictions are given in the Appendix.
Its basic assumption is that
a  stripped-envelope SNIc (a Type Ic supernova event)
results in the axial emission of opposite jets of one or more CBs, made of ordinary matter.
The observable ones have initial Lorentz factors (LFs)
$\gamma_0\!\equiv\!\gamma(t\!=\!0)$, of ${\cal O}(10^3)$.

The electrons in a CB inverse-Compton scatter (ICS) photons 
in the parent SN's ``glory".
This results in a  $\gamma$-ray beam of aperture $\simeq\! 1/\gamma_0\!\ll\! 1$
 around the CB's direction.
Viewed by an observer at an angle $\theta$ relative to the CB's direction, the  photons
are boosted in energy by a Doppler factor 
$\delta_0\!\equiv\!\delta(t\!=\!0)\!=\!1/[\gamma_0\,(1\!-\!\beta\,\cos\theta)]$
 or, to a very good
approximation for $\gamma_0^2\!\gg\!1$ and $\theta^2\!\ll\!1$:
\begin{equation}
 \delta_0=2\gamma_0/ (1+\gamma_0^2\theta^2)\, .
 \label{eq:delta}
 \end{equation}
  
CBs are efficient relativistic {\it magnetic  rackets}.
We assumed in DD2008 that the racket's ``strings'' were the magnetic fields {\it inside}
a CB. Here we assume that the racket is the (external) magnetic field of the CB \cite{LastDD}. 
As detailed in Section \ref{subsec:Revamp}
and the Appendix, the CB-model predictions stay put.
Non-solar primary cosmic-ray nuclei and electrons are made by the 
``collisions" between this field and the ambient ISM matter, previously ionized 
by the extremely intense $\gamma$-ray radiation of the GRB pulse that
a  CB made. In an ``elastic'' collision the maximum 
energy acquired by a CR (in $c\!=\! 1$ units) is:
 \begin{equation} 
 E_{\rm max}=2\,\gamma_0^2\,M\, ,
 \label{eq:Emax}
 \end{equation}
 with $M$ the mass of the CR electron or nuclear isotope.
 Spoiler: these are the energies of the spectral ``knees".
 
 It must be emphasized ab initio that, while the CB model of GRBs is used to extract
 many ``prior'' parameters compatible with independent observations
 --such as the typical properties of CBs and SN glories-- the
 same model correctly describes nuclear CRs at all energies with only one
 parameter chosen to fit the data [DD2008]. The description is a {\it first order} one
 for several reasons, e.g.:$\!$ the Galaxy's magnetic fields and the details of
 how they ``confine" CRs as a function of their momentum are not 
 precisely known, secondary CRs slightly contaminate the spectra of the
 primary ones. These are some of the {\it next order} effects that we do not address.
 
 \subsection{The unresolved conundrum}
 
 How is it possible to endow a CB with a humongous LF of ${\cal{O}}(10^3)$?
 The energy release in a SN  is of ${\cal{O}}(M_\odot /10)$, the binding energy
 of a neutron star. The mass of a CB implied by GRB data is of ${\cal{O}}(10^{-7}M_\odot)$.
 On average $\sim\!10$ of them are made in an observed GRB. Thus the energy
 carried by these CBs is ``only" of ${\cal{O}}(10^{-3}M_\odot)$.  The typical kinetic
 energy of a SN's ejecta is $> \! 50$ times larger \cite{Chang-Goo}. SN explosions are not
 yet well understood. Deciphering how all
these things  happen to happen would require a mastery of transient, catastrophic, 
relativistic, chaotic, turbulent magneto-hydrodynamics. Not  a trivial task.
``Seeing is believing'' proofs would be images of CBs traveling with enormous LFs.
Such images do exist, see the ``Superluminal Motion" entry in the Appendix.

 \subsection{An interlude on correlations}

The ICS of glory photons of energy $\epsilon$ by a CB boosts their energy, as seen by an
observer at redshift $z$, to $E_\gamma\!=\!\gamma_0\,\delta_0\,\epsilon/(1\!+\!z)$. 
Consequently, the peak energy $E_p$ of their time-integrated energy distribution satisfies
\begin{equation}
(1+z)\,E_p\!\approx\! \gamma_0\,\delta_0\, \epsilon_p\,, 
\label{eq:Ep0}
\end{equation}
with $\epsilon_p$ the peak energy of the glory's light, for which we choose
$\epsilon_p=1$ eV,  one of the CB model's priors [DD2008].

In the Thomson regime the nearly isotropic distribution (in the CB's rest frame)
of a number $n_\gamma$
of  photons is Compton scattered into an angular distribution
$dn_\gamma/d\Omega\!\approx\! (n_\gamma/4\,\pi)\,\delta^2$
in the observer's frame. Thus, the isotropic-equivalent
total energy of the photons satisfies
\begin{equation}
E_{iso}\!\propto\! \gamma_0\, \delta_0^3\, \epsilon_p. 
\label{eq:Eiso}
\end{equation}
GRBs
 --viewed from at $\theta\!\sim\!1/\gamma$,
implying $\delta_0\!\sim\!\gamma_0$-- satisfy 
\begin{equation}
(1+z)\,E_p\propto [E_{iso}]^{1/2},
\label{eq:Corr1}
\end{equation}
while far off-axis ones ($\theta^2\! \gg \! 1/\gamma_0^2$, implying $\delta_0\!\propto\! \theta^{-2}$) 
may be dubbed XRFs, have a much lower $E_{iso}$, and satisfy 
\begin{equation}
(1+z)\,E_p\propto [E_{iso}]^{1/3}.
\label{eq:Corr2}
\end{equation}

\begin{figure}[]
\vspace{.1cm}
\centering
\epsfig{file=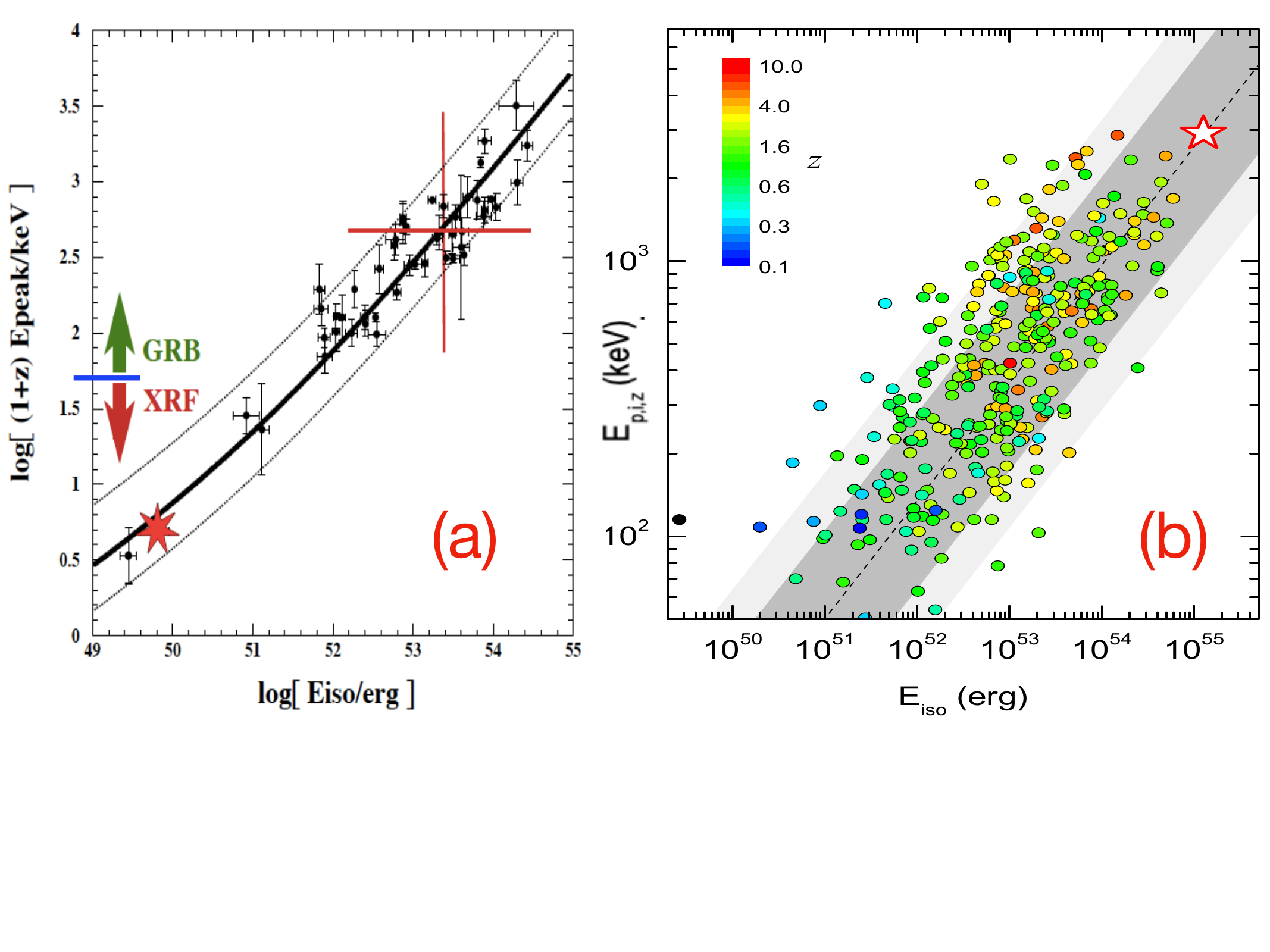,width=8.6cm}
\vspace{-1.8cm}
\caption{(a) The $[E_p, E_{iso}]$ correlation, with a slope 
interpolating the ones of Eqs.(\ref{eq:Corr1}) and (\ref{eq:Corr2}).
By definition the blue line separates GRBs from XRFs.
(b) Results for
315 long GRBs with known redshift   
observed by Konus-Wind \cite{KW}. The color of each data point represents the 
GRB's redshift. 
Error bars are not shown.
GRB221009A is indicated by a red star.
The best fit ``Amati'' correlation is plotted as a dashed line.}
\label{fig:EpEisoDouble}
\vspace{-.2cm}
\end{figure}

The predicted 
$[E_p, E_{iso}]$ 
correlations \cite{Correls} are shown in Fig.(\ref{fig:EpEisoDouble}a),
where a fit was made interpolating a power law from a $1/3$ to a $1/2$ behavior.
This ``Amati" correlation was later observationally discovered  \cite{Amati2002}. Some
recent results \cite{KW} are shown in Fig.(\ref{fig:EpEisoDouble}b), where
the best-fit  line has a slope
$0.42$, which happens to be the average of the slopes in Eqs.(\ref{eq:Corr1}) and (\ref{eq:Corr2}).

\subsection{CB-model consequences of GRB221009A}

\begin{figure}[]
\vspace{-.2cm}
\centering
\epsfig{file=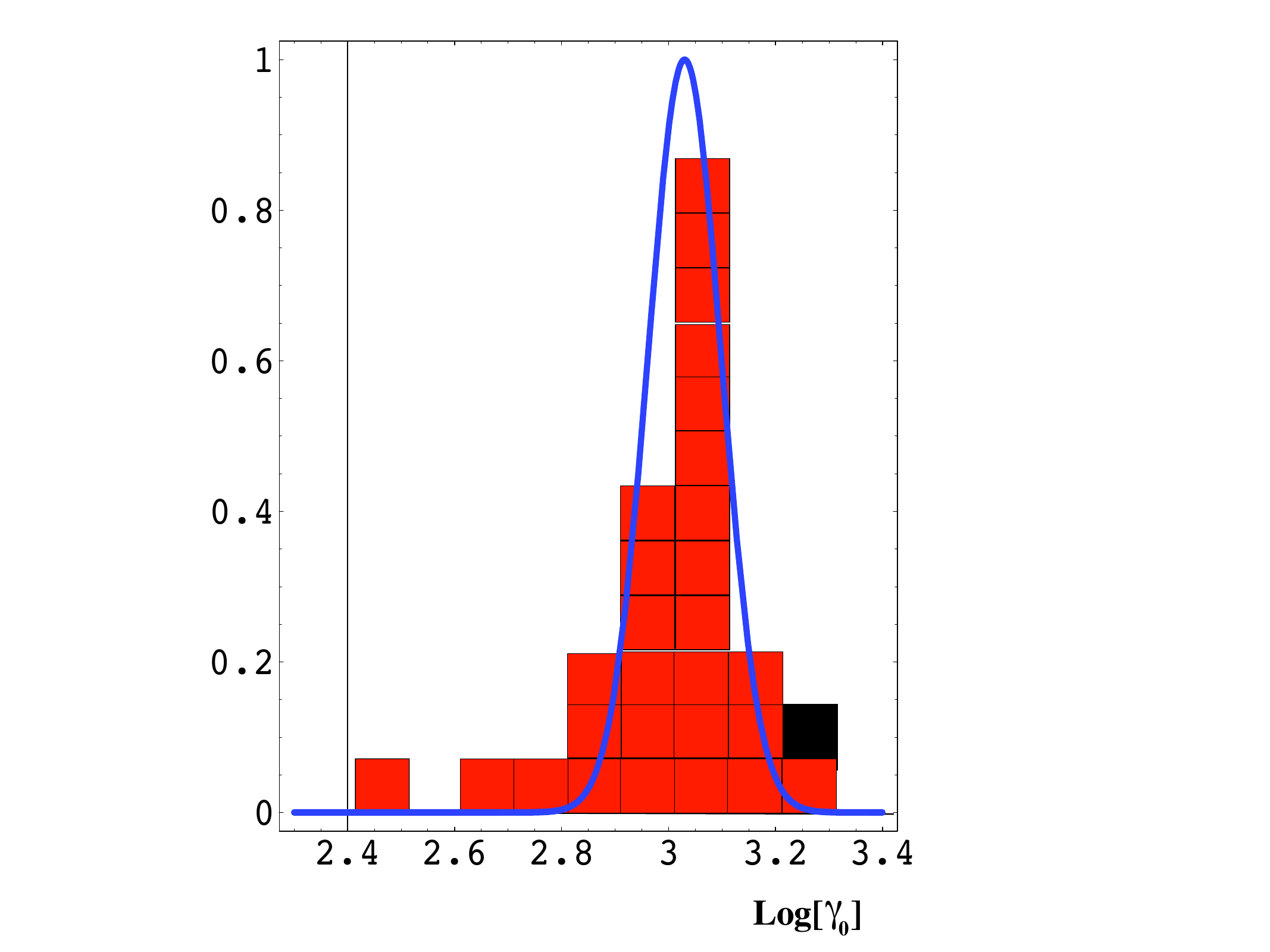,width=9cm}
\vspace{-.8cm}
\caption{Distribution of CB Lorentz factors, from the analysis of their GRB
 afterglows and a fit to it [DD2008] and, in black, the result for GRB221009A, extracted from
the peak energy of its $\gamma$ rays. Low LF CBs no doubt escape detection.}
\label{fig:GammaDist}
\end{figure}

The fact that GRB221009A, like all others, satisfies the Amati correlation is not what makes it
interesting. The implications of its measured its measured $(1+z)\,E_p=3503\pm 133$ keV, are.
Being so energetic, the CB model implies that it was observed at a small angle, so that
$\delta_0\approx\gamma_0$ and Eq.(\ref{eq:Ep0}) implies a large
$\gamma_0 [221009A]\!\approx\! 1.87 \times 10^3$.
This result is shown in black in Fig.(\ref{fig:GammaDist}), where GRB221009A is seen along
with old record breakers.

\begin{figure}[]
\vspace{-1.cm}
\hspace{-.5cm}
\centering 
\epsfig{file=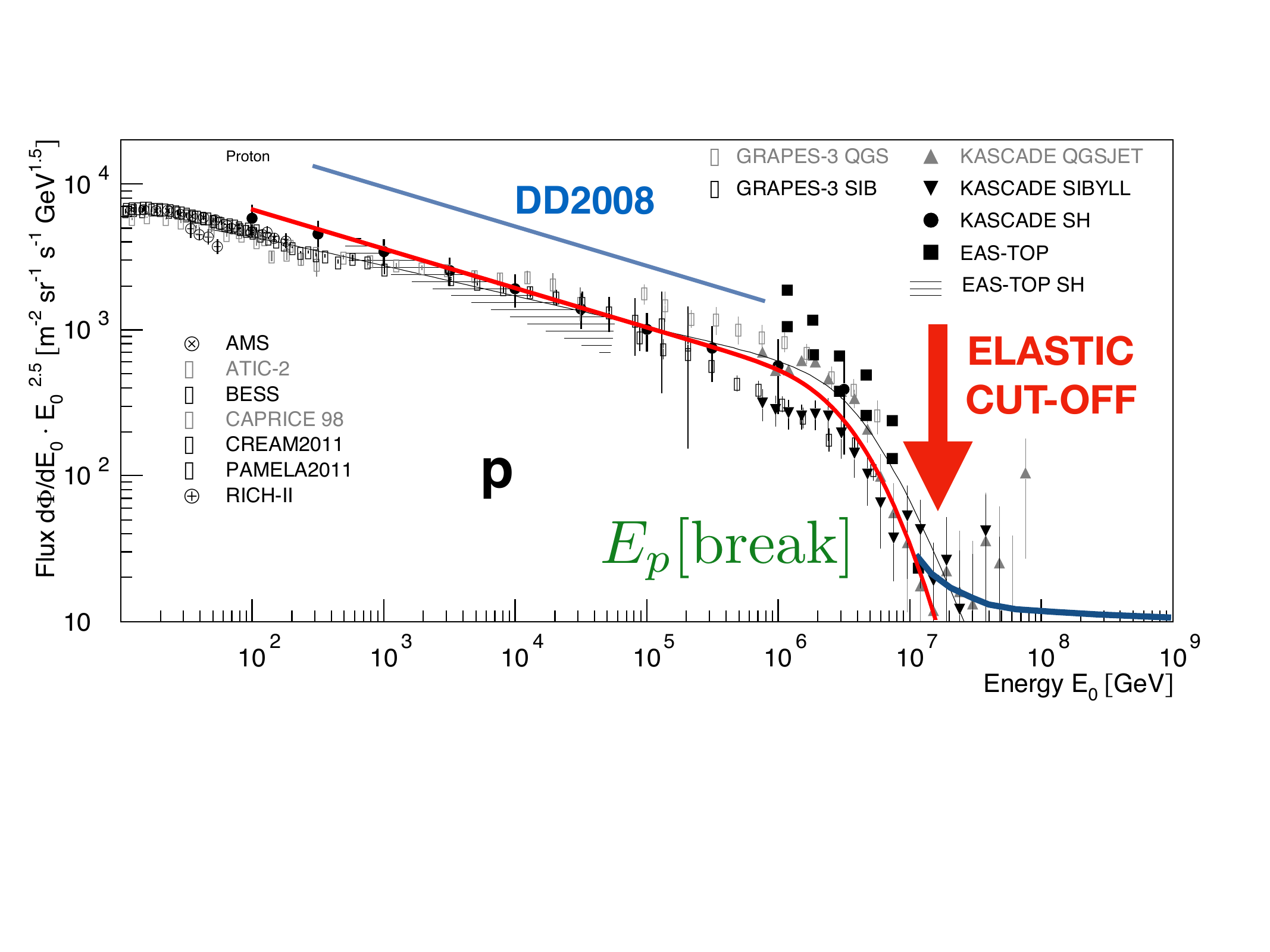,width=9.5cm}
\vspace{-2.3cm}
\caption{ The p (or ``H") knee extends from 
$E_p\!\simeq\! 10^6$ 
to $\simeq \! 10^7$ GeV.
Above the knee the  ``inelastic" contribution, discussed in Section IV, dominates.}
\vspace{-.6cm}
\label{fig:pSpectrum}
\end{figure}

A spectrum of CR protons \cite{KneeNews} is shown in Fig.(\ref{fig:pSpectrum}).
The theoretical spectrum is shown in red and the specific shape of its knee reflects the
fact that it has been predicted --and convoluted-- with the fit $\gamma_0$ distribution
shown in Fig.(\ref{fig:GammaDist}), extracted
from our analyses of  {\it GRB afterglows} (AGs) \cite{DD2004,DD2008}, not from data 
concerning CRs.

Combine Eqs.(\ref{eq:Emax}) and $\gamma_0 [221009A]\!\approx\! 1.87 \times 10^3$
to obtain a predicted
``end of the proton's knee" at an elastic cutoff energy $1.4\times 10^7$ GeV,
shown in Fig.(\ref{fig:pSpectrum}). This is the first of the a series of
good news. It verifies the idea that GRBs and CRs have a common origin.

\subsection{Criticism of the CB model}

A frequent critique of the CB model is that:
{\it The CB should swipe-up the plasma in its path, inducing a shock, compressing and 
deforming it in the process.} This is based on the traditional view that
SN shocks --or, more recently, narrow jets produced by 
SNe\footnote{If jets indeed consist in successive blobs of matter
--as implied, 
by successive GRB pulses-- that would be the skeletal CB model.}--
are the accelerators of CRs. The most convincing defense of this claim is the 
development of instabilities and shocks in numerical simulations of the collision of
two plasmas. These simulations typically employ unrealistically large electron-to-ion mass
ratios (e.g. $\! 1/15$) and relative Lorentz factors of the colliding plasmas much smaller 
than $10^3$ \cite{Sims}. Also, the simulations are one-dimensional, with periodic
boundary conditions in the transverse directions. 
It is not obvious that the results would be similar in realistic simulations of CBs.

One problem with the traditional views is that they do not result in successful comparisons 
with the data, unless they include, for instance,  acceleration by Galactic Alfv\'en waves and 
a large number of fit parameters \cite{Galprop}.

Another not infrequent critique of the CB model is: {\it It does not 
demonstrate what the dynamics of the collision between a CB and the ambient matter
would really look like.} In DD2008 we
did explicitly introduce such dynamics (Section III). The problem is that it looks so naive.

\subsection{Naive and first-principled results}

Occasionally extremely naive predictions have a success that sophisticated
first-principle derivations lack. One example is the QCD-improved 
quark model, wherein hadrons are analogous to atoms, made of constituent quarks 
whose hyperfine interactions are dictated by one gluon exchange \cite{DGG}.
This ``naive" model explained the masses of all known S-wave mesons and baryons. 
And it precisely and
correctly predicted the masses of their then unknown charmed counterparts \cite{ADR}.
It took several decades for ``first-principled" lattice gauge
theory computations to obtain similarly precise ``post-dictions'' \cite{RPP}.

A naive theory feels less naive if its predictions are abundant and successful.
The CB model of CRs makes the predictions discussed here. Its
comparison with the standard views on GRBs is  telling, see Table III
in \cite{DDDcomparisons}.

\subsection{Revamping the theory itself}
\label{subsec:Revamp}

In DD2008 we assumed that the ISM ingredients entering a CB exited
it by diffusion in its inner magnetic field and that CBs are decelerated
by ingurgitating a large fraction of the ISM they intercept\footnote{An assumption
supported by the predictions for GRB afterglows.}.
 The details of the
process --such as the CB's radius-- disappeared from the answer for  
the shape of the relativistic flux, which below the knees took the analytic form:
\begin{eqnarray}
{dF_{\rm elast}\over d\gamma}&\propto& n_{_A}
\int_1^{\gamma_0}{d\bar\gamma\over \bar\gamma^{7/3}}
\int_{\rm max[\bar\gamma,\gamma/(2\,\bar\gamma)]}
^{\rm min[\gamma_0,2\,\bar\gamma\,\gamma]}
{d\gamma_{\rm co}\over \gamma_{\rm co}^{4}}\; ,
\label{NRFlux2}
\end{eqnarray}
with $n_{_A}$ the number densities in the ISM.

CBs may have high spin and an intense magnetic field \cite{LastDD},
like newly-born neutron stars. In such a case it may be this {\it external}
magnetic field that converts the ISM it encounters into CRs. If the latter
exit the CB's ``magnetic domain'' by diffusion (as in DD2008) Eq.(\ref{NRFlux2})
would result. This ``revamped'' version of the CB model evades the
critique that magnetic fields in a CB's trajectory deviate CRs and
the CB would catch up with them \cite{Hillas,Answer}. 
A tiny fraction of ${\cal{O}}(10^{-15})$ of these ``first-injected'' CRs are Fermi-accelerated 
within the CB's magnetic domain while diffusing out of it, to result in the same flux above 
the knees as in DD2008, whose predictions are then unchanged by these revamped assumptions.
 
 First-principle simulations of a rapidly moving and rotating magnetized CB
 encountering the ISM and producing CRs would be difficult, and welcome.

\section{Universality of the CR fluxes}

For nuclear CRs, in a domain of relativistic energies below the 
corresponding knee for each $A$, the source flux of Eq.(\ref{NRFlux2}) is very
well approximated by
\begin{equation}
{dF_{s}^A/d\gamma}
\propto n_{_A}
\,\gamma^{-\beta_s},\,\,\beta_s=13/6.
\label{composource}
\end{equation}

The observed spectrum should be steeper than in Eq.(\ref{composource}), since 
Galactic confinement is a species-dependent effect, traditionally simplified
for nuclear CRs
as a power-law modification of the source flux by a confinement-time factor
$\tau\propto\left({Z/p}\right)^{\beta_{c}}$, or, in the relativistic domain,
\begin{equation}
\tau\propto\left({Z/E}\right)^{\beta_{c}}
=\left[{Z/(A\, m_n\,\gamma)}\right]^{\beta_{c}}. 
\label{galconf}
\end{equation}
All in all\footnote{An AI program may notice that we erased a factor
$(Z/A)^{\beta_{c}}$
in Eq.(\ref{eq:composource2}). This is because we assume, as in DD2008,
 that CRs exit their birthplace area by diffusion, as they do
in the Galaxy as a whole. 
This compensates exactly for the $(Z/A)^{\beta_{c}}$ factor. }
\begin{equation}
{dF^A/d\gamma}\propto n_{_{A}}\gamma^{-(\beta_s+\beta_{c})}\
\label{eq:composource2}
\end{equation}
In DD2008 we adopted a value $\beta_{c}=0.6$ \cite{CONFI}, resulting in
$\beta_s+{\beta_{\rm c}}\approx 2.77$, in good agreement with the
individual spectra and in excellent agreement with the 
 all-particle value $2.75$, recently measured by LHAASO-KM2A \cite{LHAASO-KM2A}. 
 
 \begin{figure}[]
\vspace{-.6cm}
\hspace{-.5cm}
\centering
\epsfig{file=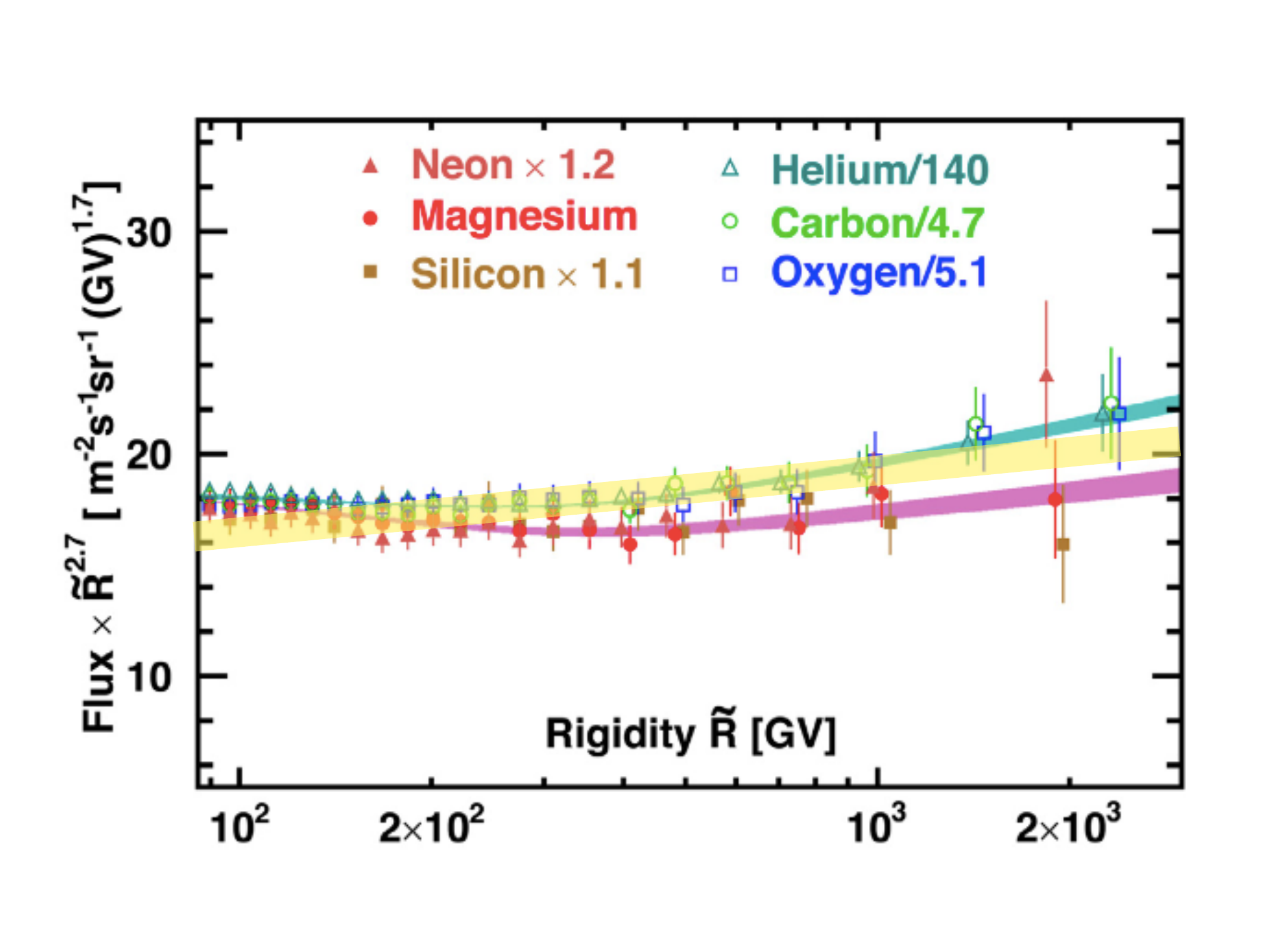,width=9cm}
\vspace{-1.3cm}
\caption{AMS results \cite{AMSresults} for various primary CRs as function of rapidity. 
The yellow band is the result in DD2008.}
\vspace{-.5cm}
\label{fig:AMSspectra}
\end{figure}

It has become customary to present results on the spectra of individual CR nuclei 
not as  functions of energy, but as functions of rigidity,
$\tilde{R}\equiv p/Z$. For the relativistic nuclei in Fig.(\ref{fig:AMSspectra}) 
$\tilde{R}\approx (A/Z)\, m_p \gamma$, or $\tilde{R}\!\propto\! \gamma$, since $A/Z$ is approximately
the same for all these nuclei. Thus the figure can be read as a test of 
Eq.(\ref{eq:composource2}).

One cannot say that Fig.(\ref{fig:AMSspectra}) is a precise
 test of the {\it predictions} in DD2008. There we chose
in $\tau$ --the CR confinement-time power law-- a value of $\beta_c$ \cite{CONFI}
compatible with the data. There is no reason for
$\tau$ to be so naively simple. Indeed, the Galaxy is complicated,
contains spiral arms, has magnetic fields whose structure is not well known, etcetera,
The observed slight hardening of the spectra at $\tilde{R}\sim 4\times 10^2\,\rm GV$
may well be due to a confinement time which is not quite a simple power law.

\section{CR elementary abundances}

We summarize here results in DD2008 because they provide a striking
test of Eq.(\ref{eq:composource2}).
In what follows we replace $\beta_s+\beta_{c}$ in Eq.~(\ref{galconf}) 
by $\beta=2.75$, the expected and/or observed all-particle spectral index.

 It is customary to present results on the composition of CRs at a fixed
energy per nucleus, $E_{_A}=1$ TeV, as opposed to a fixed $\gamma$.
Change variables
($E_{_A}\!\propto\! A\,\gamma$) in Eq.~(\ref{eq:composource2})
to obtain the prediction for the observed fluxes:
\begin{equation}
{dF_{\rm obs}/dE_{_A}}\propto \bar{n}_{_A}^{\rm amb}\,A^{\beta -1}
\,E_{_A}^{-\beta},\,\,\,\,\,\,\,\, \beta-1\sim 1.75,
\label{compo}
\end{equation}
with $\bar{n}_{_A}^{\rm amb}$ the average `ambient' ISM nuclear abundances
--listed and discussed in detail in DD2008--
in the large `metallicity' environments of the SN-rich domains wherein 
CBs produce CRs.

At  fixed energy the predictions for the CR abundances 
$X_{_{\rm CR}}(A)$ relative to protons are:
\begin{eqnarray}
X_{_{\rm CR}}(A)\approx X_{\rm amb}(A) \; A^{1.75};\;
X_{\rm amb}(A) \equiv {\bar{n}_{_A}^{\rm amb}/\bar{n}_p^{\rm amb}},
\label{composimple}
\end{eqnarray}
with $X_{_{\rm amb}}$ the abundances, relative to H (AKA p in this context), in the mentioned domains.
 \begin{figure}
\vspace{-.5cm}
\begin{center}
\epsfig{file=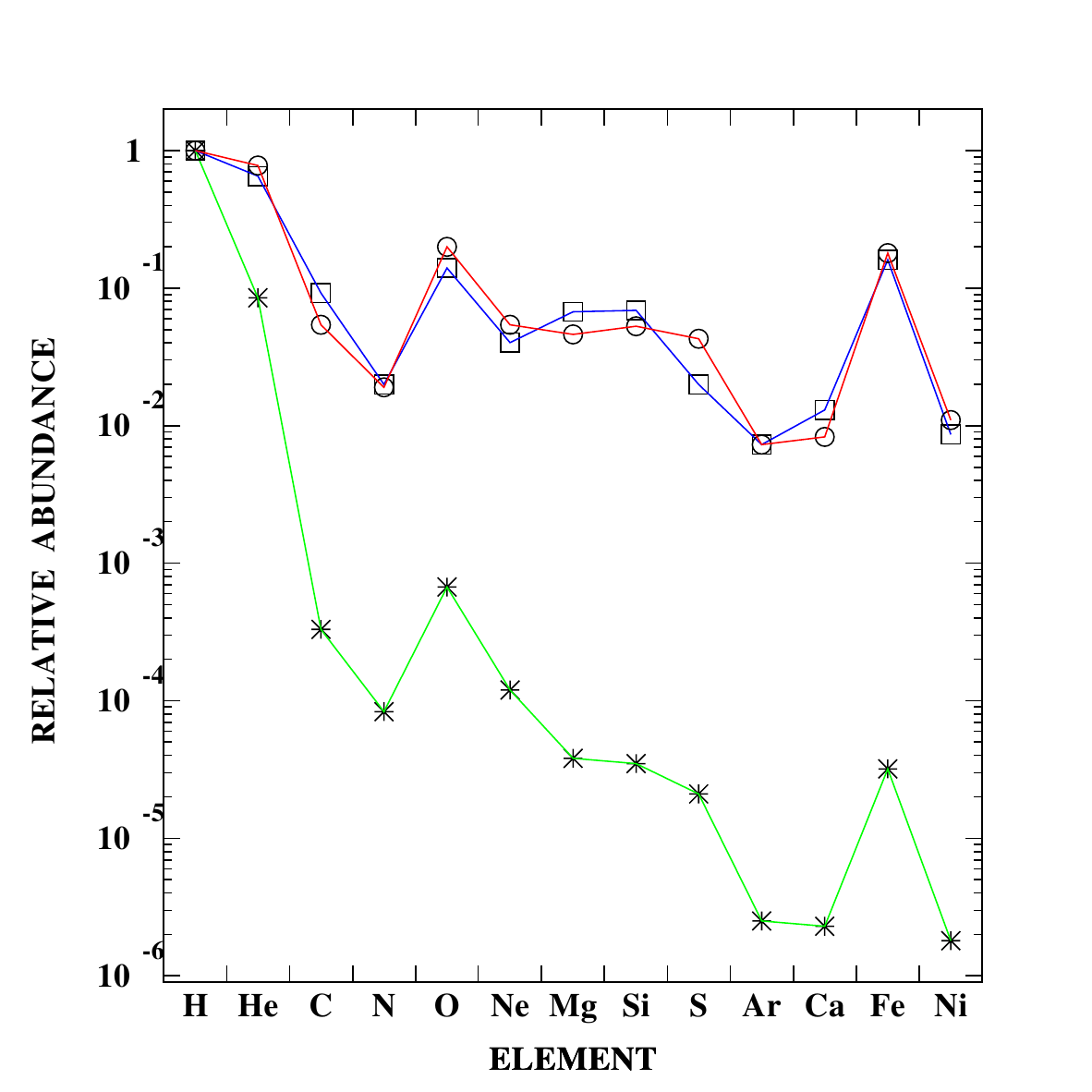, width=7cm,angle=0}
\end{center}
\vspace{-.8cm}
\caption{
The relative abundances of primary CR nuclei, from H to Ni
around 1 TeV \cite{BWS}. The
stars (joined by green lines) are solar-ISM abundances \cite{GS}.
The circles (joined in red) are the predictions, with input super-bubble
abundances. The squares (joined in black) are the CR observations.}
\vspace{-.5cm}
\label{f1}
\end{figure}

The DD2008 results are shown in Fig.(\ref{f1}). They are 
 remarkable, given that the input ambient
 abundances and the observed CR ones are quite extraordinarily
 different. The enhancement factor $A^{1.75}$ of Eq.(\ref{composimple})
snuggly serves to predict the CR abundances, and varies from 1
to $\sim\!1.22\times 10^3$ from H to $^{58}\rm Ni$. These results follow
from the simple CB-model statement that the CR fluxes, as functions
of Lorentz factor, are universal.

\section{The CR spectra above their knees}

\begin{figure}[]
\hspace{.5cm}
\epsfig{file=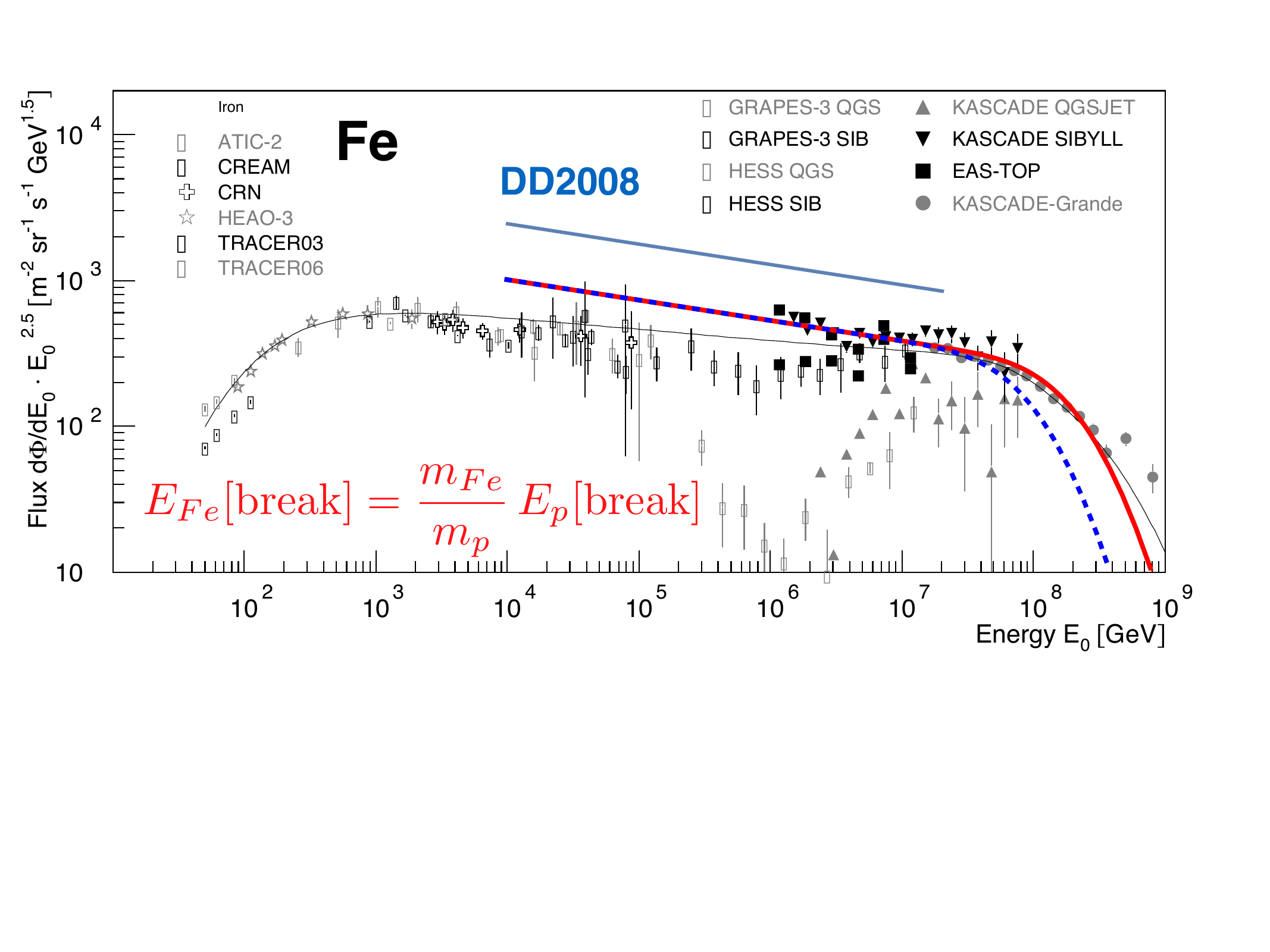,width=8.5cm}
\vspace{-2.4cm}
\caption{The Fe knee with its predicted break \cite{KneeNews}.}
\vspace{-.7cm}
\label{fig:FeSpectrum}
\end{figure}

\begin{figure}[]
\vspace{.5cm}
\hspace{-1cm}
\centering
\epsfig{file=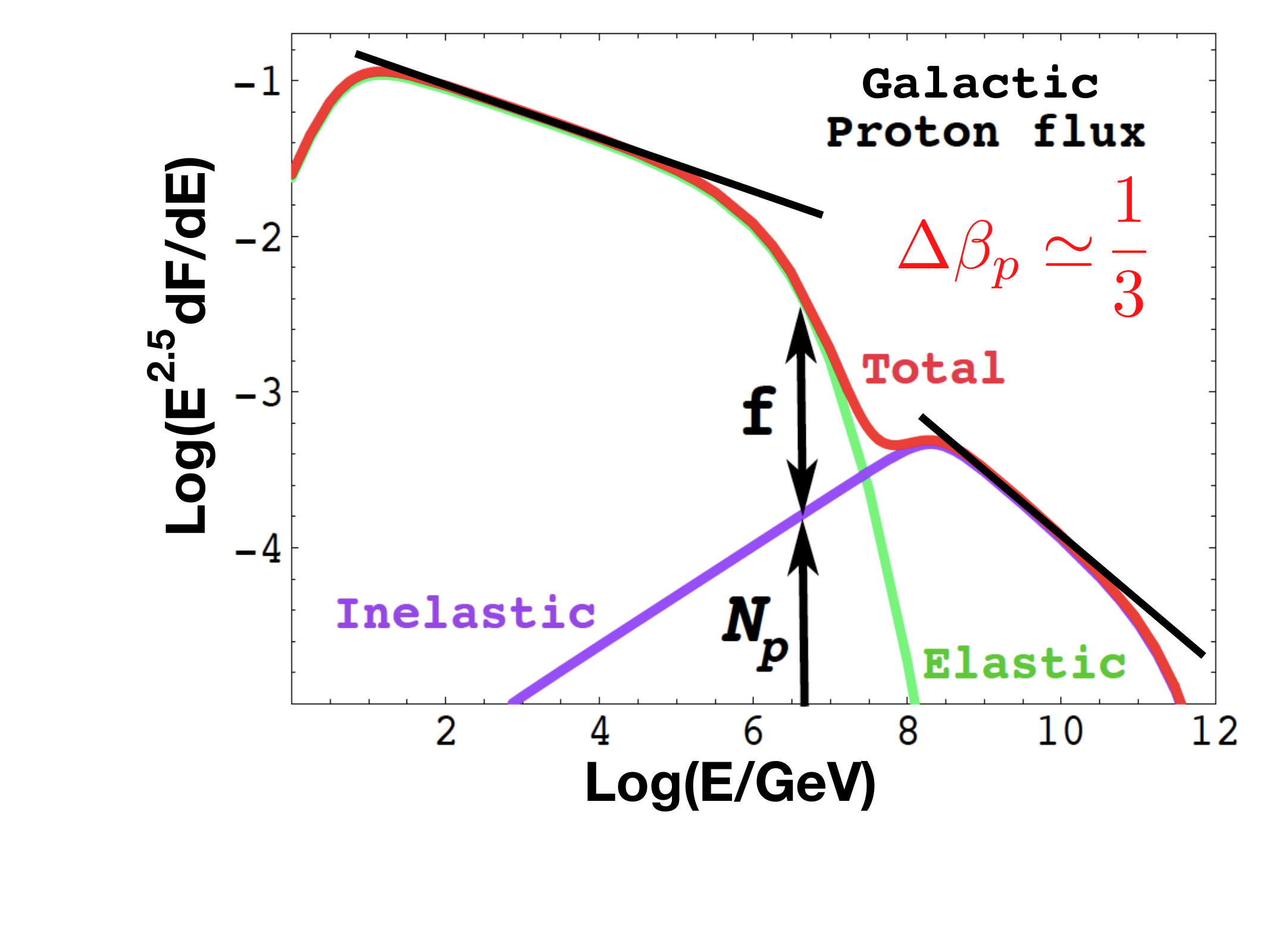,width=8cm}
\vspace{-1.cm}
\caption{The p spectrum (red), its (black) predicted slopes and
its two contributions (green and mauve) [DD2008].}
\vspace{-.5cm}
\label{fig:DD2008pSpectrum}
\end{figure}

Data of CR nuclei are sufficient to test Eqs.(\ref{eq:Emax}) and (\ref{eq:composource2}) 
for elements up to Fe, 
as shown in comparing Fig.(\ref{fig:FeSpectrum}) with Fig.(\ref{fig:pSpectrum}). But the data are insufficiently precise to test whether the knee positions scale as $A$ or $Z$. And they do not reach 
energies high enough to test the CB-model spectral predictions beyond the knees. To recall them
and to clarify a couple of points we reproduce the DD2008 prediction for the galactic flux of protons
in Fig.(\ref{fig:DD2008pSpectrum}).

Two contributions to the galactic proton flux are shown in Fig.(\ref{fig:DD2008pSpectrum}),  excluding
the extragalactic contribution, only relevant at very high energies. The  ``elastic" contribution
is the one we have discussed at length. In DD2008 we assumed that a minute fraction 
of the ISM intercepted by a CB is Fermi-accelerated by the CB's inner chaotic magnetic field, whose 
properties we deduced from a comparison to the Galaxy's magnetic fields.
When exiting the CB by diffusion these accelerated particles constituted the ``inelastic" contribution.
In the revamped CB model we assume that a similar process takes place in the ``magnetic domain"
of a CB.
Its shape is also an explicit universal function of $\gamma$, Eqs.(37,38) of DD2008.
Its predicted slope is $\simeq\! 1/3$ steeper than 
that of the elastic contribution, as shown in the figure, that defines ``slope" in the sense used here.

The quantity $f$ in Fig.(\ref{fig:DD2008pSpectrum}) describes the ratio of the two contributions at a given energy.
It cannot be fit to the proton data of Fig.(\ref{fig:pSpectrum}), but since the CB 
model correctly predicts the CR abundances of the main elements, an approximate $f$ 
can be extracted from the all-particle data above the Fe knee. Finally the combined quantity
$f\!\times\!N_p$ in the figure depends on imprecise priors, such as the rate of Galactic SNe. 
It was
adjusted in DD2008 
via the observed proton flux, related to the CR luminosity of the Galaxy, which the CB model accommodates without effort. 
The ratio of total inelastic to total elastic
fluxes is of ${\cal{O}}(10^{-15})$, re-acceleration of the initial CRs need not be very efficient.

After three consecutive paragraphs we have been unable to show the full consequences
of the prediction described in Fig.(\ref{fig:DD2008pSpectrum}). But a ``surprise" is 
awaiting in the form of the CR electron flux.

\section{the CR electron flux}

\begin{figure}[]
\vspace{-.3cm}
\hspace{-.4cm}
\centering
\epsfig{file=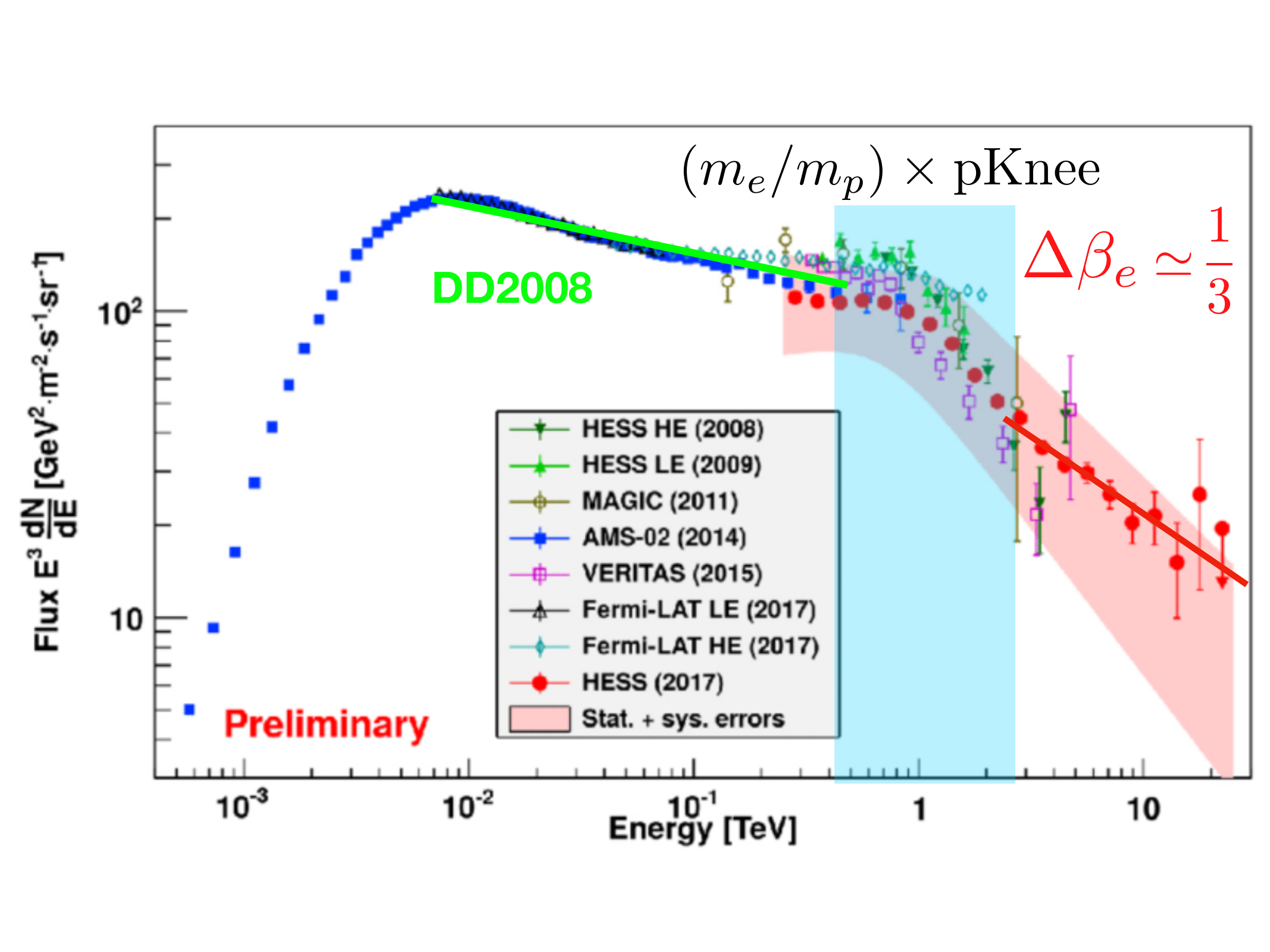,width=8.9cm}
\vspace{-1.2cm}
\caption{Data compilation \cite{Hess} and CB-model predictions for 
high energy CR electrons.
The green (red) lines are the predicted slopes below (above) the
electron knee region, symbolized by the clear blue domain which,
as predicted, occurs at an energy a factor $m_e/m_p$ times that of the
proton's knee. }
\vspace{-.5cm}
\label{fig:electronspectrum}
\end{figure}

Fig.(\ref{fig:electronspectrum}) is the electron version of the proton  result of
Fig.(\ref{fig:DD2008pSpectrum}).
The figures differ in two respects: their energies are scaled by particle
mass from a common $dF/d\gamma$ input. And, traveling in the Galaxy,
electrons loose energy faster than nuclei; interactions with magnetic
fields and the ambient radiation dominate at sufficiently high energy. A detailed
analysis in DD2008 and \cite{Positrons}
yields the prediction
 that, below the knee,
$\beta_e\!=\!\beta_s\!+\!1\!\simeq\!3.17$. Sharing a common $dF/d\gamma$ with the H spectrum,
the e$^-$ spectrum steepens as the knee is crossed by 
$\Delta\beta_e\!=\!\Delta\beta_p\!\simeq \! 1/3$. All these predictions are 
supported by the data in Fig.(\ref{fig:electronspectrum}).

The knees of CR electron, p, and other nuclear spectra and the 
peak energy of GRB221009A have a common explanation based on
the CBs' LF distribution extracted from the analysis of their afterglows. This
multiple ``coincidence" does not seem to be a random one. And
the CR break energies scale with mass, definitely not charge.

Notice that in Fig.(\ref{fig:electronspectrum}) we have not tried to deal with the
low energy data. For them the source spectrum is still a function only of their LF,
 but at low energies $v/c\!\neq\! 1$ plays a role and Eq.(\ref{NRFlux2}) is not a good
 approximation. In spite of this, in DD2008 we showed, for H and He, how the 
 CB-model successfully predicts their
observed low-energy spectral shapes, see the Appendix.
We have not made a similar
analysis of the electron lower-energy data of Fig.(\ref{fig:electronspectrum}). 
The reason
is that the task becomes arduous: for such electrons Coulomb scattering
and bremsstrahlung energy losses, 
as well as uncertainties in the confinement time of electrons, play a significant role [DD2008].

\section{The AMS positron spectrum}
\label{sec:positrons}

The CB-model's {expectation} for the spectrum of CR positrons \cite{Positrons}
is a subject that, unlike the others discussed here, cannot be summarized in simple terms.
Positrons are made in collisions of accelerated nuclei
with ambient ones, e.g.$\!$ $pp\,({\rm or} \, pn)\!\to\! \pi\,{\rm or}\, K\!\to\! \mu^+\!\to\! e^+$.

Deriving the $e^+$ flux requires a lengthy calculation involving many CB-model's inputs extracted
from other observations: the distribution on LFs shown in Fig.(\ref{fig:GammaDist}), the typical CB
baryon number ($10^{50}$), a CBs' initial expansion velocity in its rest frame
(the relativistic speed of sound, $c/\sqrt{3}$) and the SN wind's surface density
($10^{16}\rm g/cm^2$). Other inputs are the number of CR generating SNe
in the Galaxy (1/100y), and the number of CBs per SN (10, twice the average
number of GRB pulses).

\begin{figure}[]
\vspace{-.5cm}
\hspace{-.5cm}
\centering
\epsfig{file=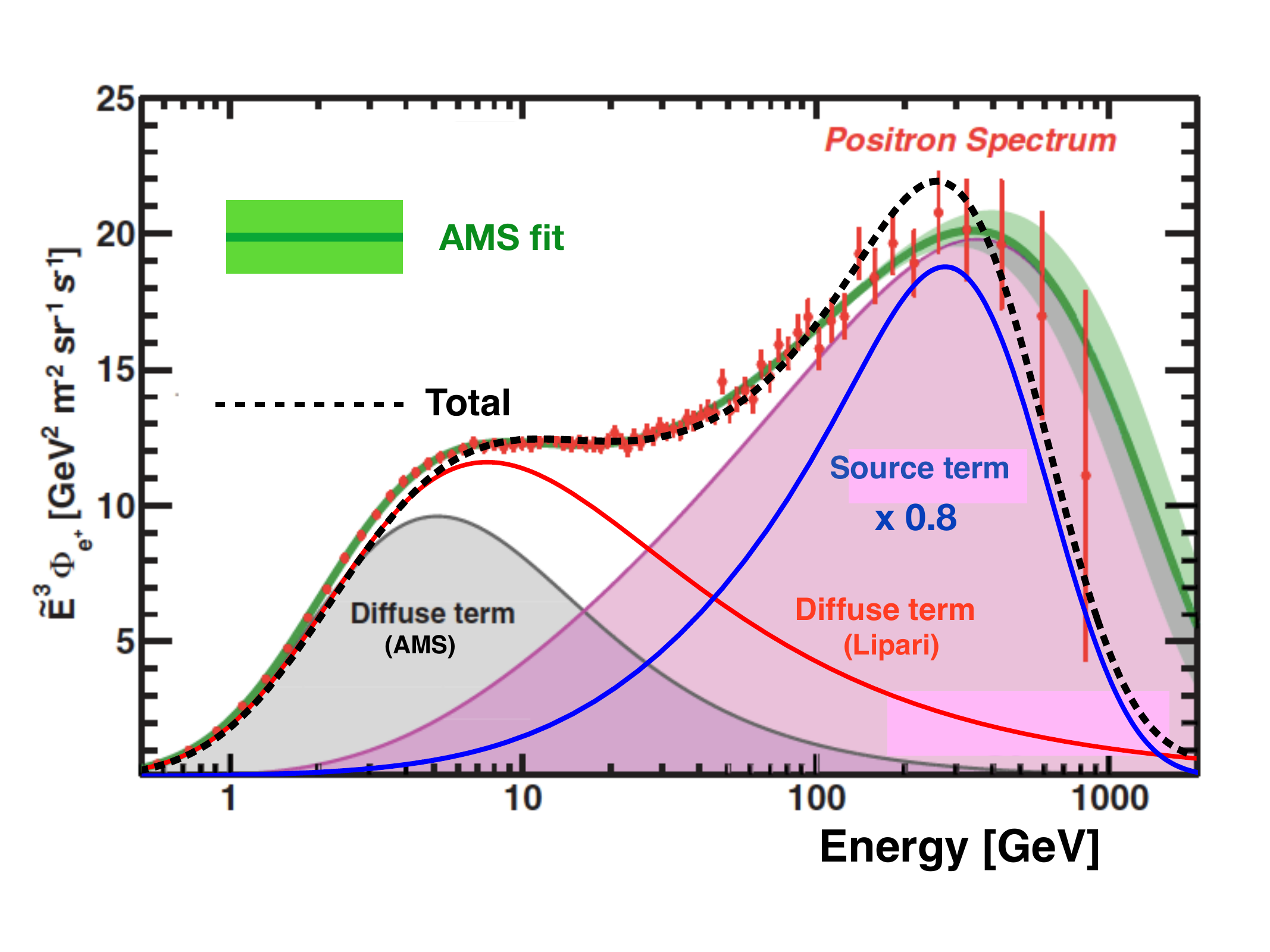,width=9cm}
\vspace{-1.3cm}
\caption{Adopted diffuse term (red) and calculated source term (blue), and their sum
(black, dashed).}
\label{fig:Results}
\end{figure}

The result of the above calculation \cite{Positrons} is shown in Fig.(\ref{fig:Results}) as
the dashed line. Some comments: secondary positrons (the ``diffuse term" in AMS
parlance) are a  not well known ``background'', we chose the one labeled 
Lipari \cite{Lipari}. For esthetics, the ``source term'' CB model's prediction has been reduced
 to 0.8 its calculated value, this fudge factor being much less important than the uncertainties
 in the inputs\footnote{As shown in \cite{Positrons}, Shlomo Dado made a perfect fit to
 the data by slightly adjusting the input priors. But that is  {\it not} the point.
 The point is that the ``source term'' in Fig.(\ref{fig:Results}) is a prediction.}.

\section{high-energy neutrinos}

\begin{figure}[]
\centering
\epsfig{file=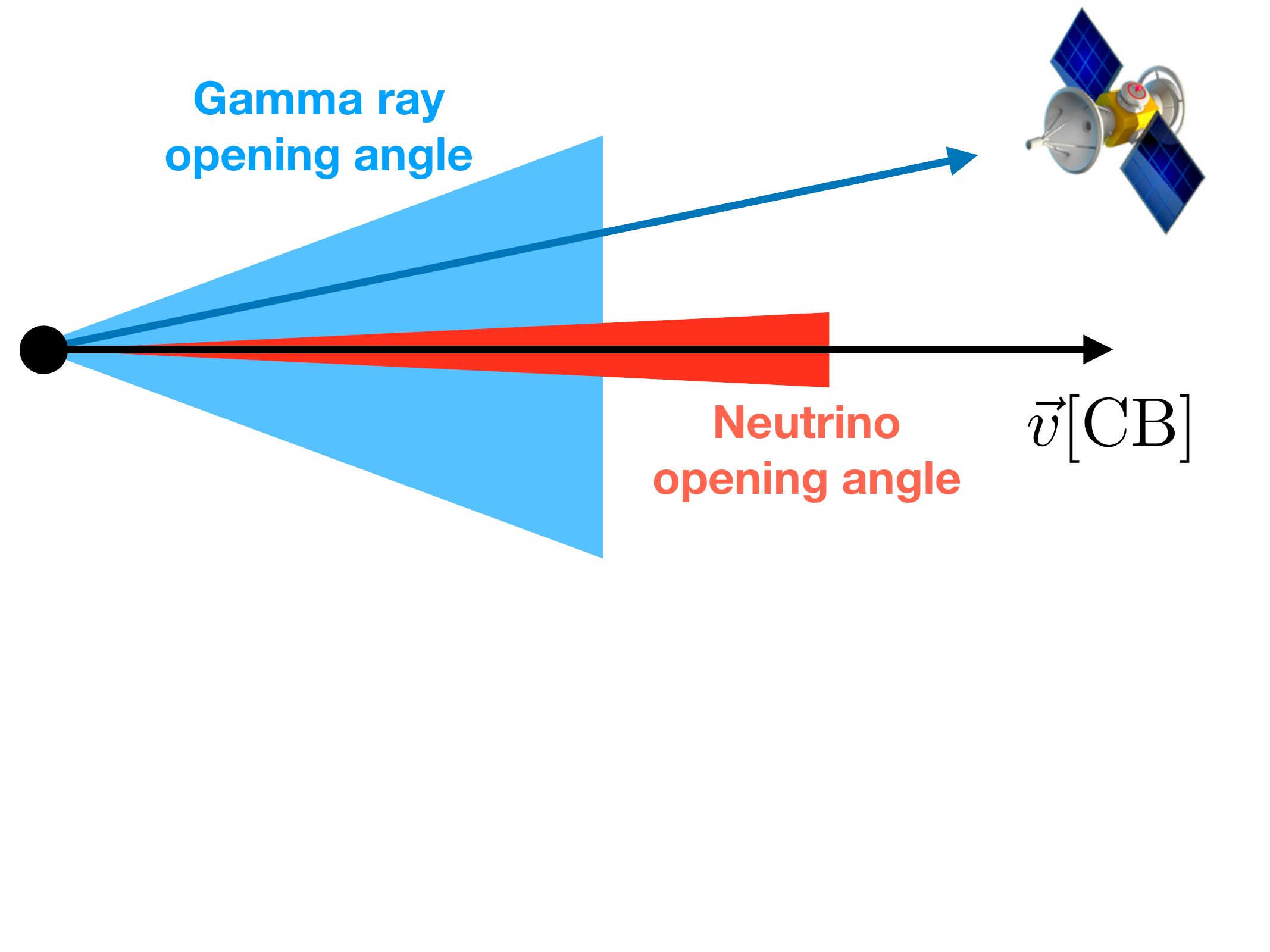,width=8.5cm}
\vspace{-2.7cm}
\caption{An observed GRB and its missing neutrinos.}
\label{fig:Neutrinos}
\end{figure}

The same processes producing positrons produce neutrinos, via
decays such as $\pi^\pm\to \mu^\pm\,\nu$ and $\mu\to e\nu\bar{\nu}$.
In the line of sight to a GRB, should neutrinos not be observable in
coincidence with the GRB proper? Why have they
not been detected by IceCube? \cite{IceCube}. The CB-model answer is deceptively
simple, see Fig.(\ref{fig:Neutrinos}). As remarked in \cite{LastDD, YoNeutrinos}
neutrinos produced in proton-induced collisions have transverse
momenta of the order of that of their parent mesons, in turn of
the order of the mesons' masses.  The consequent neutrino-beam opening
angle is much smaller than the one of observable GRB gammas.
And the GRB afterglow's neutrino flux is negligible [DD2008]. 
Moreover, to expect a neutrino signal it must be
assumed that ambient magnetic fields are
weak enough not to deviate the charged parents of neutrinos.

The electrons accelerated by a CB to $\gamma_e\!\gg\! \gamma_0$[CB]
may also scatter ambient light
to produce photons of much higher energy than the usual 
$E_\gamma$ of $\cal O$(1) MeV. But their beam has an aperture
$1/\gamma_e\! \ll \! 1/\gamma_0$[CB], explaining why there are so few
GRBs with very high energy photons.

\section{Conclusions}
\label{sec:conc}

\begin{figure}
\begin{center}
\epsfig{file=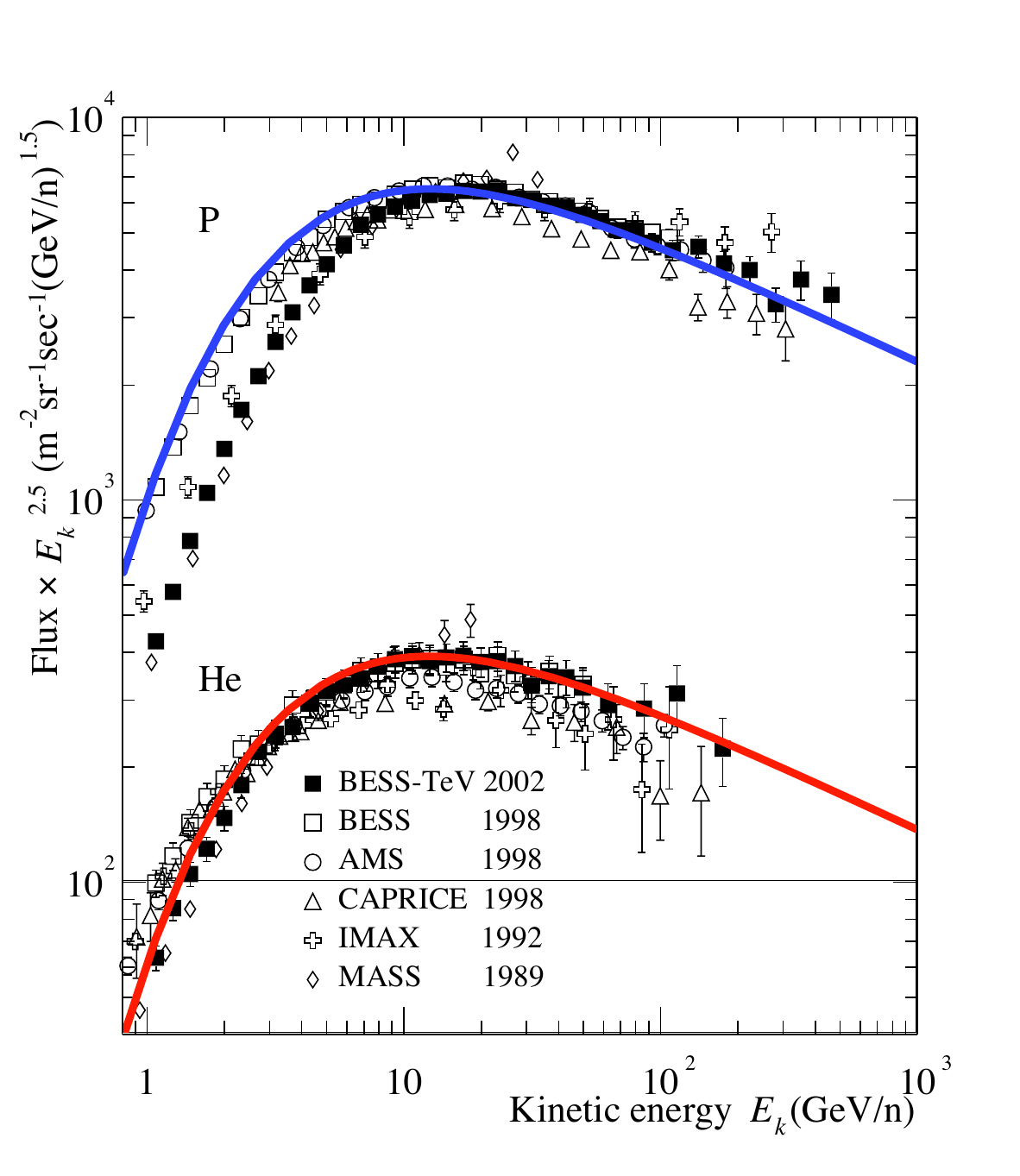, width=7.2cm,angle=-0}
\end{center}
\vspace{-.8cm}
\caption{The very low energy fluxes of protons and $\alpha$ particles
at various times in a solar cycle. The 1998 data are close to solar-minimum
time. }
\label{VeryLowEnergy}
\end{figure}

In the Appendix and in previous articles (e.g.$\,$DD2008) we have argued that 
the CB model provides a satisfactory comprehension of cosmic rays at all energies. 
 Particularly gratifying, in our view, were the analytical results on CR fluxes,
including the predictions for the low-energy CR spectra as in Fig.(\ref{VeryLowEnergy})    
and the very simple explanation of the observed relative CR abundances,
as in Fig.(\ref{f1}). These understandings are still ``first order" in that we cannot
interpret the data in minute detail, see Fig.(\ref{fig:AMSspectra}).

The CB model is unsatisfactory in the sense that, though cannonballs
have been observed moving superluminally in the plane of the sky
\cite{Taylor, GWSHB},
we do not have a
detailed analytical or numerical understanding of the interactions
between CBs and their hypothetical magnetic fields with the ISM.
But the current ``guesses" --so very successful
in the understanding
of GRBs \cite{DD2004,{DDDcomparisons}}--  continue to yield good results
for CRs as well:

The knee energies of the CR spectra of electrons, H, He and Fe nuclei scale with
mass. This is particularly striking when comparing electrons with nuclei, see
Figs.(\ref{fig:pSpectrum},\ref{fig:FeSpectrum},\ref{fig:electronspectrum}).
The H knee ends at the energy predicted via the Lorentz factor
of GRB221009A, obtained from the CB-model's interpretation of its peak energy.
That peak energy is at the top of the distribution of LFs extracted from the
CB-model analysis of GRBs, not of CRs, see Fig.(\ref{fig:GammaDist}). A
calculation --with no fit parameters-- of the positron spectrum observed by
AMS gives a convincing result, see Fig.(\ref{fig:Results}). The non observation
of very high energy neutrinos by IceCube has a trivial explanation \cite{LastDD,YoNeutrinos}. 

One might be tempted to state that, once again, Baron Munchausen's cannonballs 
 hit their various targets.

\vspace{.6 cm}
\noindent {\bf Acknowledgment:} 
I am particularly indebted to Shlomo Dado, Arnon Dar and Fabio Truc for discussions and advice.
This project has received funding/support from the European Union's Horizon 2020 research and innovation programme under the Marie Sklodowska-Curie grant agreement No 860881-HIDDeN.

\vspace{2.4 cm}
{\bf  Appendix: The CB model of GRBs and CRs}

\begin{figure}[]
\centering
\epsfig{file=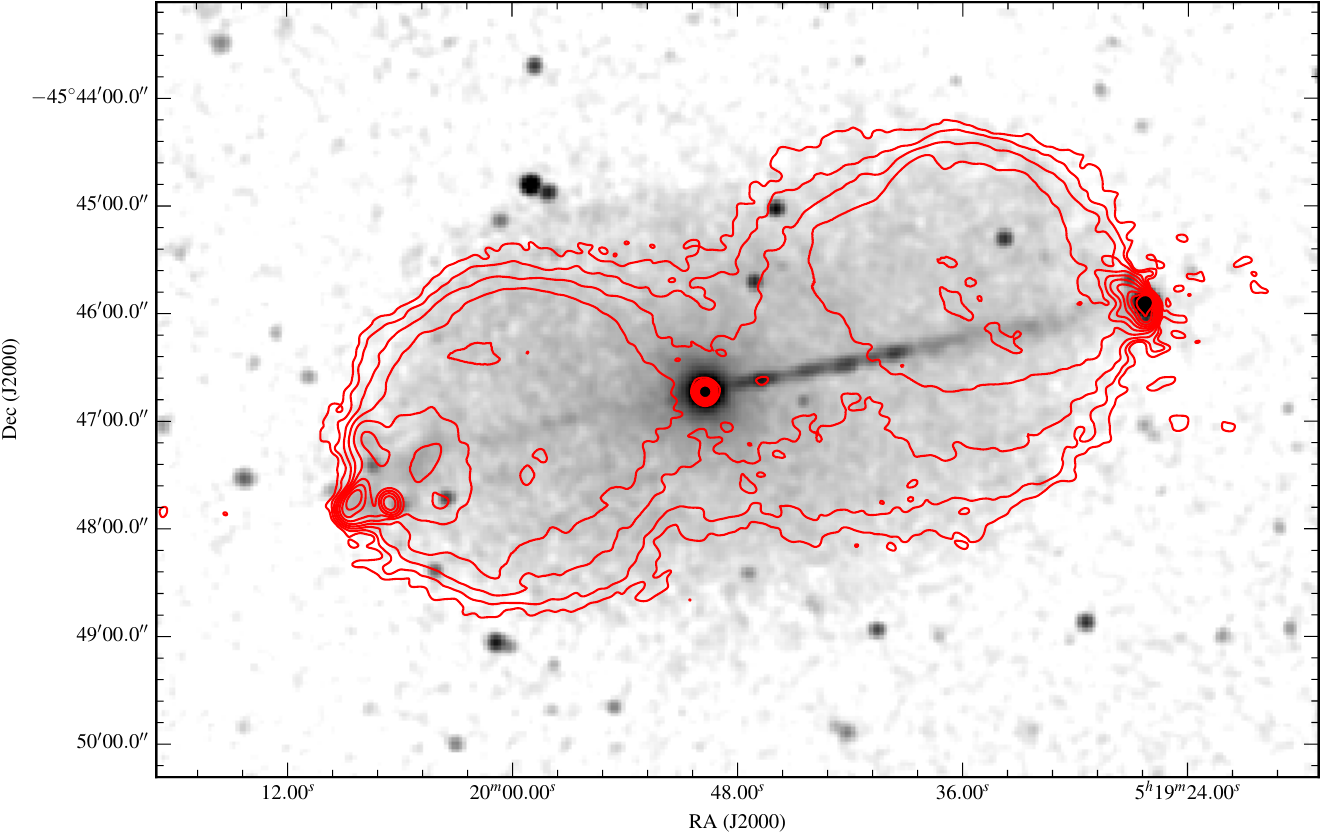,width=8.cm}
\vspace{-.3cm}
\caption{The quasar Pictor A. A superposition of an X-ray image and the (red)
contours of the radio emission
\cite{Pictor}.}
\vspace{-.5cm}
\label{fig:PictorA}
\end{figure}

Jets are emitted by many astrophysical systems, such as Pictor A,
shown in Fig.~(\ref{fig:PictorA}). Its active
galactic nucleus is discontinuously spitting something that, seen in X-rays, does not
appear to expand sideways before it stops and blows up, having by then
travelled almost $10^6$ light years. Many such 
systems have been observed. The Lorentz factors 
of their ejecta are of ${\cal{O}}(10)$.
The mechanism responsible for the ejections,
due to episodes of violent accretion into a black
 hole, is not well understood.

The radio signal in Fig.~(\ref{fig:PictorA}) is the synchrotron radiation of
`cosmic-ray' electrons \cite{Pictor}. Electrons and nuclei were scattered
by the CBs of Pictor A, which encountered them at rest in the
intergalactic medium, kicking them up to high energies.
Thereafter, these particles diffuse in the ambient
magnetic fields --that they contribute to generate-- and the electrons radiate.

In our galaxy there are `micro-quasars', whose central black
hole's mass is a few $M_\odot$. 
The best studied one \cite{GRS}
is GRS 1915+105.
A-periodically
it emits two
oppositely directed {\it cannonballs}, traveling at $v\sim 0.92\, c$.
When this happens, the continuous X-ray emissions
 ---attributed to an unstable accretion disk--- temporarily decrease.
Atomic lines of many elements have been seen
in the CBs of $\mu$-quasar SS 433 \cite{SS433}. 
Thus these ejecta are made of ordinary matter.

The `cannon' of the CB model is analogous to the ones
of quasars and $\mu$-quasars, though with larger LFs.
In the {core-collapse} responsible for a stripped-envelope
SNIc event, due to the parent star's
rotation, an accretion disk  is produced around
the newly-born compact object  by stellar material originally
close to the imploding core or by more distant matter
falling back after the shock's passage. 
 A CB made of {\it ordinary-matter plasma} is emitted, as
in $\mu$-quasars, when part of the accretion disk
falls onto the compact object. {\it Long-duration} GRBs, XRFs
and {\it non-solar} CRs are produced by these jetted CBs.

\begin{figure}
\centering
\epsfig{file=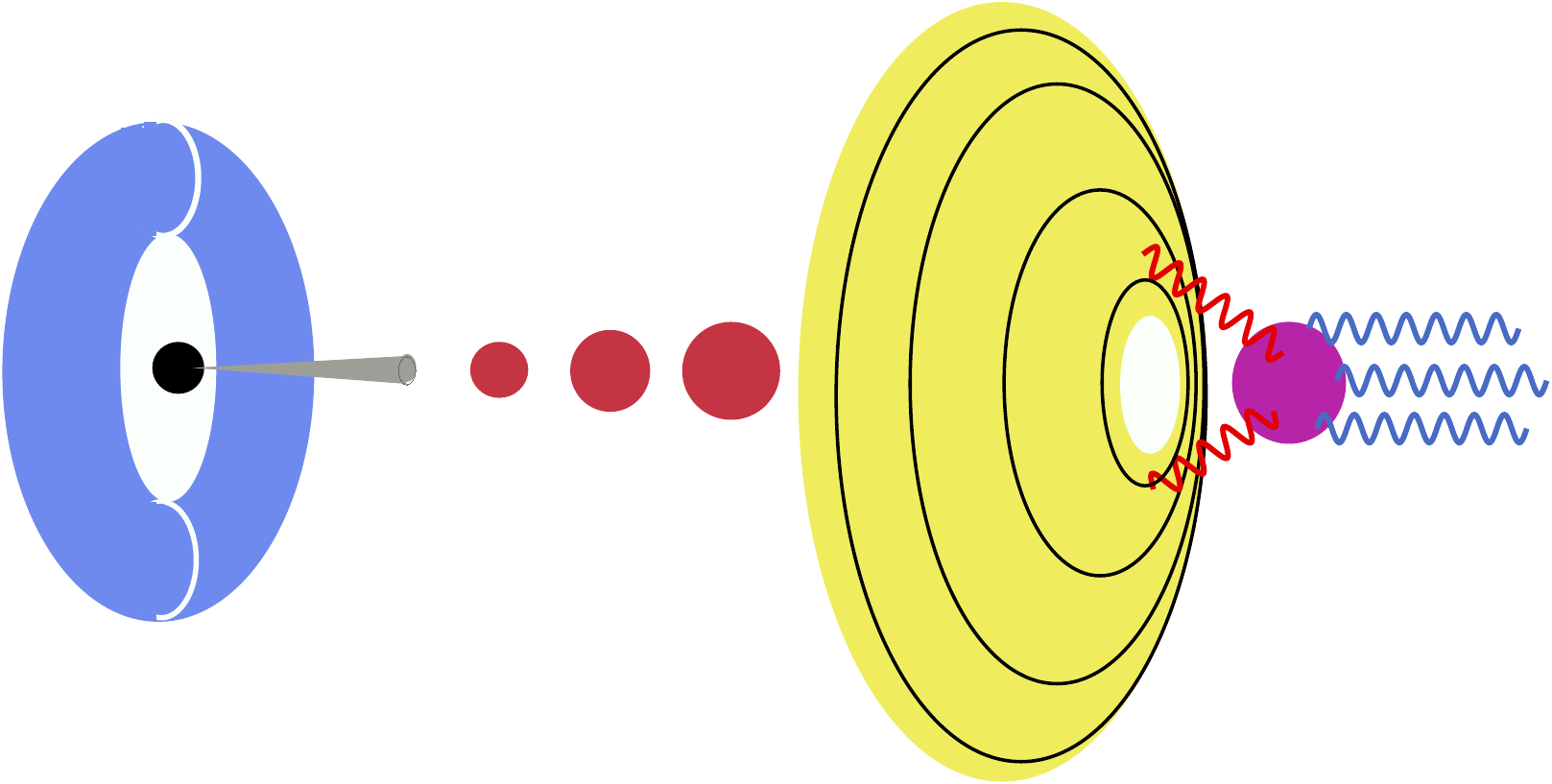,width=0.5\linewidth,angle=-90}
\caption{An `artist's view' (not to scale) of the CB model
of long-duration GRBs. A core-collapse SN results in
a compact object and a fast-rotating torus of non-ejected
fallen-back material. Matter (not shown) abruptly accreting
into the central object produces
a narrowly collimated beam of CBs, of which only some of
the `northern' ones are depicted. As these CBs move through
the `ambient light' surrounding the star, they Compton up-scatter
its photons to GRB energies.}
\vspace{-.5cm}
\label{figCB}
\end{figure}

A summary of the CB model is given in Fig.~\ref{figCB}. 
The {\it `inverse' Compton scattering} (ICS) of light by electrons within a CB  
produces a highly forward-collimated beam of higher-energy photons.
The target light is in a temporary reservoir: the {\it glory}, 
an ``echo" (or ambient)
light from the SN, permeating the ``wind-fed" circumburst density profile, previously ionized
by the early UV flash accompanying a SN explosion and/or  by the enhanced UV emission
that precedes it.
To agree with 
observations, CBs
must have baryon numbers  
$N_{_{\rm B}}$ of ${\cal{O}}(10^{50})$,
$\sim\!1/2$ the mass of Mercury, a miserable $\sim\!10^{-7}\,M_\odot$.

The simple kinematics describing a beam of GRB or XRF photons --viewed at different
angles-- suffice to predict all observed correlations between pairs of prompt 
observables, e.g.~photon
fluence, energy fluence, peak intensity and luminosity, photon energy at peak 
intensity or luminosity, and pulse duration. The correlations are tightly obeyed,
indicating that observable GRBs are moderately standard candles  --with ``absolute" properties
varying over a couple of orders of magnitude-- while the observer's angle makes
their apparent properties vary over very many orders of magnitude  \cite{corr}. 
Double and triple correlations of GRB observables and the ``break time"
of their afterglows agree with the CB model \cite{corr1}.
The shapes of GRB pulses and their spectrum are also neatly explained by 
ICS of glory light \cite{DD2004}.

In its journey through its host galaxy, a CB encounters
the constituents of the ISM, previously ionized by the GRB's $\gamma$-rays.
The CB itself or, more likely, the collision of its magnetic field with
the ISM electrons and nuclei results in CRs.
GRBs and XRFs have long-lasting {\it `afterglows'} (AGs). 
 The CB model accounts
for them as synchrotron radiation from the ambient electrons swept in 
by the CBs, predicting the correct fluencies, AG light curves and spectra
\cite{AGoptical, DDX}.

The obstacles still separating the CB model from a complete theory of CRs and GRBs are
the theoretical understanding of the CBs' ejection mechanism in SN explosions
and of the precise way in which a CB's assumed magnetic field interacts with the ISM.
Otherwise the CB model describes all known properties of GRBs and XRFs.
But, perhaps more significantly, the model also resulted in remarkable predictions:

\vspace{1.2cm}
{\bf The SN-GRB association}
\vspace{.1 cm}

GRB980425 was {\it `associated'} with SN1998bw: within directional 
errors and within a timing uncertainty of $\sim\!1$ day, they coincided. The 
luminosity of a 1998bw-like SN peaks at $\sim\!15\,(1+z)$ days. 
{\it Iff} one has a predictive theory
of AGs, one may test whether GRBs are associated with  `standard torch' SNe, 
akin to SN1998bw, `transported' to the GRBs' redshifts. 
The test was already conclusive (to us) in 2001 \cite{AGoptical} or 2002,
as evidenced by the titles of the works cited in \cite{SNdreams}, unsuccessful attempts
to convince observers to search for SNe and CBs.

One could even {\it foretell  the date} in which a GRB's SN 
would be discovered. For example, GRB030329 was so
`near' ($z\!=\!0.168$) that we could not resist posting such a daring
prediction \cite{SN030329} during the first few days of AG observations. 
The spectrum of this SN snugly coincided with
that of SN1998bw. The prediction of where and when a SN would be seen
was shortened from thousands of years to just one night.

\vspace{.2cm}
{\bf The AG light curves}
\vspace{.1 cm}

Swift has established a ``canonical behaviour'' of the X-ray and optical AGs of a 
large fraction of GRBs. The X-ray fluence decreases very fast from a
`prompt' maximum. It subsequently turns into a `plateau'. After a time of
${\cal{O}}(1$d), the fluence bends (has an achromatic `break', in the usual parlance)
and steepens to a power-decline. 
Although all this was considered a surprise, it was not \cite{canonical}. 
Even GRB980425, the first to be clearly
associated with a SN, sketched a canonical X-ray light curve, 
with what we called a `plateau' \cite{AGoptical}.
Scores of X-ray and optical AGs are  correctly
described by the CB model \cite{AGoptical, DDX}.

\vspace{.2cm}
{\bf The (apparent!) superluminal motion}
\label{SUPLUM}
\vspace{.1 cm}

One may state that to support the CB model cannonballs ought
to be clearly ``seen'', as in the $\mu$-quasar XTE J1550-564 \cite{Corbel}.
Only in two SN explosions that took place close enough the
CBs were in practice clearly observable.  One case
was SN1987A, in the LMC,
whose approaching and receding CBs were
photographed \cite{Costas}.
The other was SN2003dh, associated with GRB030329,
at $z=0.1685$. In the CB model interpretation,
its two approaching CBs were first `seen', and fit,
as the two-peak GRB 
and the two-shoulder AG.
This allowed us 
to estimate \cite{SLum030329} the time-varying angle of their apparent superluminal
motion \cite{Courdec}
in the sky. The two sources or `components' were seen in radio observations at a date coincident with an optical AG re-brightening. 

The AG light curves showed various such rebrightenings, analogous 
 to the ones of XTE J1550-564 \cite{Corbel} and due to unpredicatable density 
 inhomogeneities in the circumburst domain. Once these inhomogeneities are taken into
 account \cite{SLum030329} the data snugly agree with the predictions, including 
the  inter-CB separation \cite{SLum030329}\footnote{The
size of a CB is small enough for its radio image to
scintillate, arguably more than observed \cite{Taylor}.
But the ISM electrons a CB
scatters, synchrotron-radiating in the ambient magnetic field, 
significantly contribute at radio frequencies,  blurring the 
CBs' radio image \cite{SLum030329}. Also, during the  time 
of a radio observation the CBs move, 
obliterating the scintillations \cite{NS}.}.
The observers claimed the contrary, though the 
evidence for the weaker `second component' is $>20\sigma$,
and they \cite{Taylor} closed the issue 
 by stating in their abstract: {\it The presence of this component is not expected from the 
standard model.}\footnote{Imagine the reaction to a similar challenge in cosmology
or particle physics, realms where there are ``standard" models strongly supported
by data and successful predictions.}

The proof that, contrary to (2003) standard views, GRBs are generated by Type Ic SNe and
their observed sources are superluminal CBs induced the community to react fast and Urbi et Orbi. 

The New York Times ``NYT030620'' reported  \cite{Overbye}
{\it Astronomers described the results as the ``smoking gun'' they had long suspected must be there connecting the two most violent phenomena in nature,} GRBs and SNe (not quasars, in their view). Also, the ``suspicion" of a SN/GRB connection was not held before by the people the NYT consulted, there had been no reaction to \cite{SNdreams}.

\begin{figure}
\vspace {-1cm}
\begin{center}
\epsfig{file=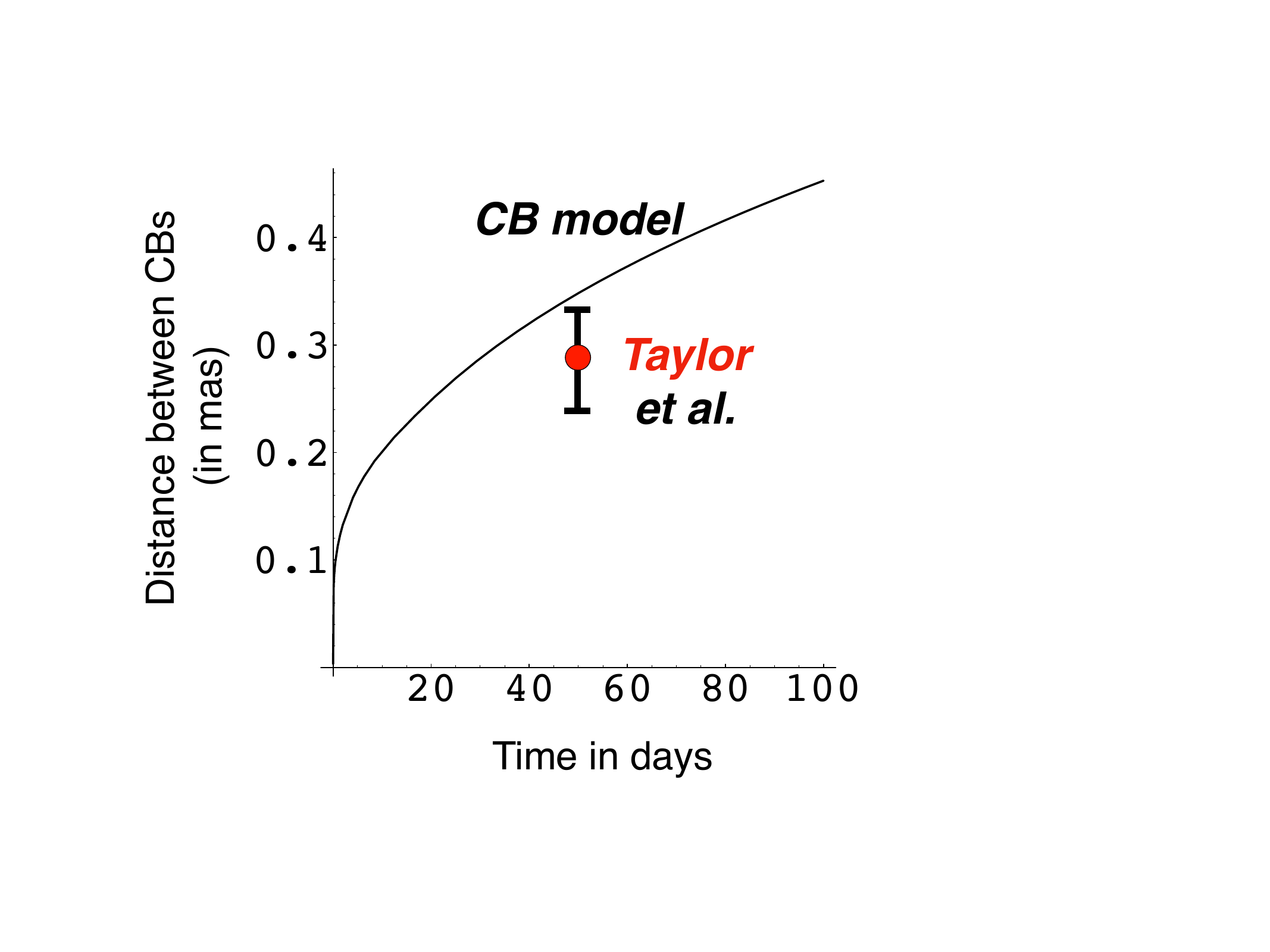, width=11cm,angle=-0}
\end{center}
\vspace{-2cm}
\caption{The expected CB-model angular distance between the
two CBs \cite{SLum030329} and the observed one \cite{Taylor}.}
\label{fig:CBangular}
\end{figure}

Even earlier, in ``NYT030529" \cite{NYT} 
one learned that {\it Dr. Dale A. Frail... said: The findings not only supported the standard model,} 
{\it but are also ``sufficient to rule out predictions of the cannonball model.''} The first half of this quote is not surprising, the second may have been a bit exaggerated\footnote{An early version of \cite{JSB}
stated that {\it Owing to the
proximity and bright radio emission, high-resolution  VLBA
imaging of the compact afterglow was used by Frail (2003) to {\bf unequivocally disprove} the
cannonball model for the origin of GRBs.}}. 
Indeed, as shown in Fig.(\ref{fig:CBangular}), theory and observation disagreed, but by less than $2\sigma$.

***

The event GW170817/SHB170817A is not a long, but a short GRB: a neutron star merger
and gravitational-wave source. In its radio afterglow, 
 a CB was observed with an overwhelming statistical significance 
 ($>\!17 \sigma$)
and traveling in the plane of the sky, as expected in the CB model, at a apparent velocity 
$V_{\rm{app}}\sim 4\,\rm c$ \cite{GWSHB}.

 {\bf The Low-Energy CRs}

 The universal flux of Eq.(\ref{eq:composource2}) is a $\gamma\!\gg\!1$
 approximation of a slightly more complicated expression, Eq.(33) of DD2008,
 applicable for all CR energies below the individual 
 knees. For p and He fluxes, the result is shown in
 Fig.(\ref{VeryLowEnergy}). A remarkable point is that in the figure only
 the absolute flux normalizations --and no other parameters-- have been
 adjusted within the predicted uncertainties.

\vspace{.2 cm}

\vspace{2cm}
 {\bf The GRB's $\gamma$-ray polarization}\newline\newline
Earliest but not least \cite{SD,DDDpol}. Let a CB launched with a LF $\gamma_0$ be seen at an angle $\theta$ 
from its jetted direction. The observed $\gamma$-rays, having been Compton up-scattered,
have a polarization $\Pi\!\approx\! 2\,\gamma_0^2\,\theta^2/(1+\gamma_0^4\,\theta^4)$.
This vanishes on axis, is nearly 100\% for the most probable viewing angle ($\theta\!\sim\! 1/\gamma_0$) and
$>\! 47$\% for $2/\gamma_0\!>\!\theta>1/(2\,\gamma_0)$. All measured GRB polarizations \cite{pol1}
are $>\!47$\%, but two, 930131 and 100826A, whose polarizations are
also incompatible with $\Pi=0$, the standard expectation for synchrotron radiation of electrons
in a non ad-hoc-structured magnetic field.

\end{document}

\bibitem{Cocconi}%
G. Cocconi, Nuovo Cimento, {\bf 3}, 1433 (1956).

\bibitem{Morrison}%
P. Morrison, Rev. Mod. Phys. {\bf 29},  235 (1957).

\bibitem{Auger}
The Auger Collaboration, Science {\bf 357},  1266 (2017); arXiv:1709.07321v1.

\bibitem{GZK}%
K. Greisen, Phys. Rev. Lett. {\bf 16}, 748, (1966);
G.T. Zatsepin \& V.A. Kuzmin, JETP Lett. {\bf 4}, 78 (1966).

\bibitem{DP}%
A. Dar \& R. Plaga, Astron. \& Astrophys. {\bf 349}, 259 (1999);
 A. Dar, A. De R\'ujula \& N. Antoniou, Proc. Vulcano Workshop 1999
(eds. F. Giovanelli and G. Mannocchi) p. 51
Italian Physical Society, Bologna-Italy, astro-ph/9901004;
A. De R\'ujula, Nucl. Phys. Proc. Suppl. {\bf 151}  23 (2006);
 hep-ph/0412094;

\bibitem{DD2008}
A. Dar and A. De  R\'ujula,
 Phys. Rept. {\bf 466}, 179 (2008), hep-ph/0606199. 
 
 \bibitem{SD}
 N. J. Shaviv and A. Dar, ApJ {\bf 447}, 863 (1995).
 
 \bibitem{GRB1}%
A. Dar \& A. De R\'ujula, astro-ph/0008474.

\bibitem{DD}%
A. Dar \& A. De R\'ujula, Phys. Rept. {\bf 405}, 203,  (2004).

 \bibitem{AGoptical}%
S. Dado, A. Dar \& A. De R\'ujula,  Astron. \& Astrophys. {\bf 388}, 
1079 (2002).

\bibitem{AGradio}%
S. Dado, A. Dar \&  A. De R\'ujula, Astron. \& Astrophys. {\bf 401},  
243 (2003).

 \bibitem{DDDXRF}%
S. Dado, A. Dar \&  A. De R\'ujula, Astron. \&Astrophys. {\bf 422},  
381 (2004).

\bibitem{DDGBR} %
A. Dar \&  A. De R\'ujula, Mon. Not. Roy. Astr. Soc. {\bf 323},  
391 (2001); 
S. Dado, A. Dar \& A. De R\'ujula, Nuc. Phys.  {\bf B165}, 103 (2007).

 \bibitem{CDD}%
S. Colafrancesco, A. Dar \&  A. De R\'ujula, Astron. \& Astrophys. {\bf 413},
441 (2004).

\bibitem{NS}
S. Dado, A. Dar \& A. De R\'ujula. arXiv:1712.09970.

\bibitem{Hor}
J.R. Hoerandel, Proc.ÒWorkshop on Physics of the End
of the Galactic Cosmic Ray SpectrumÓ, Aspen, USA,
2005, astro-ph/0508014.

\bibitem{AMS}
M. Aguilar et al., Phys. Rev. Lett. {\bf 117}, 231102 (2016).

\bibitem{JSM}
See, for instance, G. J\'ohannesson et al., arXiv:1602.02243 and references therein.

\bibitem{BWS}
 B. Wiebel-Sooth, P. Bierman \& H. Meyer, Astron. \& Astrophys. {\bf 330}, 37 (1998).
 
 \bibitem{GS}
N. Grevesse \& A.J. Sauval, 1998, Sp. Sci. Rev. {\bf 85}, 161
(1998); N. Grevesse \& A.J. Sauval, Adv. Sp. Res. {\bf 30}, 3,
(2002).

\bibitem{Hor2}
J.R. Hoerandel, Invited talk, Centenary Symposium 2012: Discovery of Cosmic Rays, June 2012, Denver.
AIP Conf.Proc. {\bf 1516}, p52 (2012).
arXiv:1212.0739v1.

\bibitem{DaDa}
S. Dado \& A. Dar, arXiv:1504.03261.

\bibitem{rpp}
J.J. Beatty, J. Matthews, \& S.P. Wakely in Review of Particle Properties.\\
http://pdg.lbl.gov/2017/reviews/rpp2017-rev-cosmic-rays.pdf

\bibitem{Dampe}
DAMPE Collaboration, Nature {\bf 552}, 63 (2017).
arXiv:1711.10981v1, and references therein.

\bibitem{AMS2}
L. Accardo et al., Phys. Rev. Letters {\bf 113}, 121101 (2014).

\bibitem{CALET}
O. Adriani et al., Phys. Rev. Letters {\bf 120}, 261102 (2018); astro-ph/1806.09728.

\bibitem{ADRprecise}
A. De R\'ujula, arXiv:0711.0970.

\bibitem{Pictor}
M. J. Hardcastle et al. MNRAS, {\bf 455}, 3526 (2016).

\bibitem{GRS}
I.F. Mirabel \& L.F. Rodriguez, Annu. Rev. Astron. Astrophys.
37, 409 (1999).

\bibitem{SS433}
S.S. Eikenberry et al., Astrophys. J. 561, 1027 (2001);
D.R. Gies et al., Astrophys. J. {\bf 566}, 1069 (2001);
M.G. Watson et al., Mon. Not. Roy. Astron. Soc. {\bf 222}, 261 (1986); 
T. Kotani et al., Publ. Astron. Soc. Jap. {\bf 48}, 619 (1996); 
H.L. Marshall, C.R. Canizares \& N.S. Schulz, Astrophys. J. {\bf 564}, 941 (2002); 
S. Migliari, R. Fender \& M. M\'endez, Science {\bf 297}, 1673 (2002); 
M. Namiki et al., Publ. Astron. Soc. Jap. {\bf 55}, 1 (2003).

\bibitem{corr}
A. Dar \&  A. De R\'ujula, arXiv:astro-ph/0012227;
S. Dado, A. Dar \&  A. De R\'ujula, Astroph. J. {\bf 663}, 400, (2007),
Astroph. J. {\bf 693}, 311, (2009), Astroph. J. {\bf 696}, 994, (2009).

\bibitem{corr1}
S. Dado \& A. Dar, Astroph. J. {\bf 755}, 16, (2013).

\bibitem{merger}
J.K. Frederiksen et al., Astrophys. J. 608, L13 (2004).

\bibitem{DDX}
S. Dado \& A. Dar, Phys. Rev. {\bf D94},  063007 (2016).

\bibitem{SN030329}
S. Dado, A. Dar \&  A. De R\'ujula, Astroph. J. {\bf 594}, L89 (2003).

\bibitem{canonical}
S. Dado, A. Dar \& A. De R\'ujula, Astroph. J. {\bf 646}, L21  (2006).

\bibitem{Costas}
P. Nisenson \& C. Papaliolios,  Astrophys. J. {\bf 518}, L29, (1999).

\bibitem{SLum030329}
S. Dado, A. Dar \&  A. De R\'ujula, arXiv:astro-ph/0402374, 0406325

\bibitem{Taylor}
G.B.~Taylor {\it et al}.~ Astrophys.~J.~{\bf 609}, L1, (2004).

\bibitem{DDDpol}
S. Dado, A. Dar \&  A. De R\'ujula, arXiv:astro-ph/0403015; arXiv:astro-ph/0701294 in
Proceedings of the Vulcano Workshop on Frontiere Objects in Astrophysics, May 21-27, 2006, 
Vulcano Italy.

\bibitem{pol1}
GRB 021206. W. Coburn \&  S. E. Boggs, Nature {\bf 423}, 415 (2003).\\
GRBs 930131, 960924. D. R. Willis,  
et al. 2005, A\&A, {\bf 439}, 245 (2005).\\
GRB 041219A: E. Kalemci et al.
ApJS, {\bf 169}, 75 (2007).\\
GRB 041219A: McGlyn et al.  
A\&A, {\bf 466}, 895 (2007).\\
GRB 100826A: Yonetoku et al. ApJ, {\bf 743}, L30 (2011).\\
GRBs 110301A and 110721A: Yonetoku et al.  ApJ, {\bf 758}, L1 (2012).\\
GRB 061122: Gotz et al. MNRAS {\bf 431}, 3550, (2013).\\
GRB 140206A: Gotz et al. MNRAS {\bf 444}, 2776 (2014).

\end{thebibliography}
\end{document}